\documentstyle[preprint,prb,aps]{revtex}
\begin{document}
\draft

\title{Magnetic phase diagram and transport properties 
of ${\rm \bf FeGe_{2}}$}
\author{C.P. Adams, T.E. Mason, S.A.M. Mentink, and E. Fawcett}
\address{Department of Physics, University of Toronto, Toronto, Ontario,
Canada M5S 1A7\\}
\date{\today}

\maketitle

\begin{abstract}
We have used resistivity measurements to study the magnetic phase diagram of 
the itinerant antiferromagnet ${\rm FeGe_{2}}$ 
in the temperature range from 0.3--$300 \: {\rm K}$ in magnetic fields up 
to 16 T. 
In contrast to theoretical
predictions, the incommensurate spin density wave phase is found to be 
stable at least up to ${\rm 16 \: T}$, with an estimated critical field 
$\mu _{\rm \/ 0}H_{\rm \/ c}$ of $\sim $ 30 T\@.
We have also studied the low temperature
magnetoresistance in the [100], [110], and [001] 
directions.  The transverse magnetoresistance  
is well described by a power law for magnetic fields above 1 T with no
saturation observed at high fields.
We discuss our results in terms of the magnetic structure and 
the calculated electronic bandstructure of ${\rm FeGe_{2}}$.
We have also observed, for the first time in this compound, 
Shubnikov-de Haas oscillations in
the transverse magnetoresistance with a frequency of 190 $\pm $ 10 T for
a magnetic field along [001].
\end{abstract}


\section{Introduction}
There is growing evidence that the proximity of the high ${\rm T_{c}}$
oxides to a spin density wave (SDW) instability is responsible for their 
anomalous normal state transport properties\cite{takagi89} and possibly 
for the superconductivity 
itself.\cite{scalapino87,monthoux92,levin94,ruvalds95}
This has led to a renewed interest in systems which exhibit similar 
instabilities since there are many unanswered questions regarding the
transport and magnetic properties of SDW systems.  The 
prototypical SDW system is the itinerant antiferromagnet Cr and dilute
alloys of Cr with other transition metals.\cite{fawcett94} 
The tetragonal intermetallic compound ${\rm FeGe_{2}}$ belongs to this 
class of materials and, in its
undoped form, exhibits both commensurate and incommensurate magnetic
ordering as temperature is varied.
SDW ordering also occurs in Mott-Hubbard systems, such 
as ${\rm V_{2-y}O_{3}}$, where strong correlations
lead to a breakdown of conventional band theory and, as in the high 
${\rm T_{c}}$ cuprates, a metal-insulator transition.\cite{bao93}

The properties of ${\rm FeGe_{2}}$ have been studied 
since 1943 and many of the details of the
crystal and magnetic structure have been
determined.\cite{walbaum43,corliss85,dorofeyev87,menshikov88}
${\rm FeGe_{2}}$ is
characterized by two magnetic phase transitions, the first of which is the
N\'{e}el transition from a paramagnetic phase to an incommensurate SDW 
state at $T_{\rm \/N}$ = 289 K 
in zero field.  The second one is a
commensurate-incommensurate (C-IC) transition into
a collinear antiferromagnetic structure.  This 
transition occurs at
$T_{\rm \/c}$ = 263 K in zero field.  
The propagation vector $\vec {Q}$ of the SDW is parallel to [100]
and varies
from 1 to roughly 1.05 $a^{*}$ between the two transitions
which are seen in
resistivity\cite{krentsis70,krentsis73}, ultrasonic 
attenuation\cite{pluznikov82},
heat capacity\cite{corliss85,mikhelson71}, and AC
susceptibility\cite{tarasenko89} studies.  A direct determination of the
change in magnetic structure has been performed by elastic neutron 
scattering.\cite{corliss85,dorofeyev87,menshikov88}
Theoretical predictions have been put forward as to the nature
of the magnetic phase diagram\cite{menshikov288,zainullina89} 
and the magnetic parameters associated with it.\cite{piratinskaya79} 
Although this compound has been well characterized in zero field there 
has been very little work done in magnetic fields and not much 
is known about the details of the magnetic phase diagram.

This work explores the magnetic phase diagram and provides an 
experimental test of the theoretical predictions.
We have used resistivity measurements along the [110] crystal direction 
in single crystal ${\rm FeGe_{2}}$ 
to find the
phase transitions and determine the field and temperature
dependences for fields up to 16 T along [110] and [100].
We find that the C-IC phase boundary rises steeply with
magnetic field and would not intersect the N\'{e}el transition boundary
until ${\mu _{\rm \/ 0}H_{\rm \/ c} \simeq }$ 30 T for a simple extrapolation.
This is in sharp contrast to 
theoretical predictions of ${\mu _{\rm \/ 0}H_{\rm \/ c} \sim }$
1 T.\cite{tarasenko89}  We have also
performed a detailed study of the low-temperature 
magnetoresistance of ${\rm FeGe_{2}}$.
The results reveal a moderate anisotropy between the c-axis and 
the basal plane and large
differences in the longitudinal and transverse magnetoresistance.
We have also observed Shubnikov-de Haas
oscillations for a magnetic field along [001] at
a frequency of $190 \pm 10 \: ~
{\rm T}$ which had not been seen in  previous ${\rm de Haas}$-
${\rm van Alpen}$
experiments.\cite{svechkarev80}

${\rm FeGe_{2}}$ has a tetragonal ${\rm CuAl_{2}}$ structure with 
symmetry group
${D_{\rm 4h}^{\rm 18}}$, space group $I4/mcm$.  Its crystal  
structure is shown
in Fig. 1.  Notice that each layer of germanium 
atoms is a mirror
image in [110] of its nearest-neighbor layers (a glide transformation).  
All germanium atoms lie in planes normal to [001] between 
the planes of iron atoms.
The length of the a-axis is 5.908 {\AA } and the length of the 
c-axis is 4.955 {\AA } at room temperature.
Hence, the distance between iron atoms is 2.478 {\AA } 
along the c-axis and 4.178
{\AA }  in the (001) plane.  The distance in the [001] direction is very
close to the distance in elemental iron (2.86 {\AA }).  
It is likely that 
the primary magnetic interaction along the
c-axis is due to overlap of {\em d}-orbitals, as it is
in elemental iron, whereas the magnetic interaction in the 
plane is mediated by superexchange and the conduction electrons,
leading to a net antiferromagnetic interaction.
The low temperature structure has ferromagnetic
alignment of iron atoms along the c-axis chains
and antiferromagnetic alignment between the chains, resulting in a 
collinear antiferromagnetic structure which is formed from two magnetic
sublattices. The incommensurate SDW phase is characterized by four Bragg
peaks displaced from the commensurate (100) position along the (100) and
(010) directions similar to spin ordering observed in 
${\rm La_{2}NiO_{4+\delta}}$\cite{tranquada94} and 
${\rm La_{2-x}Sr_{x}NiO_{4}}$\cite{hayden92} although in the latter case the
correlations are short ranged.  In the doped ${\rm La_{2}NiO_{4}}$ the
SDW ordering is coupled to charge density wave ordering (CDW) reflecting the
strong coupling between charge and spin degrees of freedom in the
transition metal oxides.

\section{Experimental Results and Discussion}
Single crystals of ${\rm FeGe_{2}}$ were cut by spark erosion from the same 
sample employed in neutron scattering studies.\cite{holden94}
The samples were less than 1 mm in thickness and width and
between 5 mm and 7 mm long.  The resistance measurements were made using an AC
resistance bridge with a frequency of 16 Hz\@.  A 
platinum resistance thermometer was used for all of the high
temperature measurements and a carbon-glass thermometer was used for
the low temperature, high magnetic field measurements.  The resistivity ratio 
(${\equiv \rho \, {\rm (300 \: K)/}
\rho \, {\rm (0.3 \: K)}}$) was 50 for all directions.

The zero-field
resistivity along [100], [110], and [001], measured between 0.3 K and 300 K,
is shown in Fig. 2.  Notice that there is a factor of two
anisotropy between the basal plane measurements and those along the
c-axis and only
moderate anisotropy in the basal plane itself.  
This is indicative of
the difference between the atomic structure along the 
c-axis iron
chains and the in-plane directions.  The resistivity along the
c-axis is remarkably linear between about 70 K and room temperature.  The two
phase transitions are well resolved as anomalies in the [110] data as
shown by arrows in the inset to Fig. 2.  
The transition temperatures are $T_{\rm \/ c}=$
263 ${\pm }$ 2 K and $T_{\rm \/ N}=$ 289 ${\pm }$ 2 K, in good
agreement with those measured by neutron diffraction.\cite{corliss85}
The [100] data shows the transitions in a similar way.  The data for 
current along [001], where the transitions are subtle,
were not used in construction of the phase diagram.  This 
difference between the signatures of the two transitions for 
in and out-of plane directions is not surprising since the propagation vector
of the SDW and the direction of the magnetic moments lie in the basal
plane, not along the c-axis.
Error associated with the transition temperature values arises from
uncertainty in locating the phase transition since the anomaly occurs
over a rather large range and we do not have a theoretical prediction
for $\rho (T)$ near the phase transitions where the phonon background is
large.
We define ${T_{\rm \/ c}}$
by the knee of the $\rho (T)$ curve 
(the lower transition) and  ${T_{\rm \/ N}}$ by the inflection point 
of the $\rho (T)$ 
curve (the upper
transition) yielding the same transition temperatures
as reported by Corliss {\em et al. }\cite{corliss85}
Our data is in fair 
agreement with that of Krentsis {\em et al.}
\cite{krentsis70}  
The compared values of room temperature resistivity are 
$167 \: \pm \: 17 \: \mu \Omega {\rm -cm}$ and 
$140 \: \pm \:  5 \: \mu \Omega {\rm -cm}$ for 
the c-axis and 
$72   \: \pm \: 8   \: \mu \Omega {\rm -cm}$ and 
$80 \: \pm \: 3 \: \mu \Omega {\rm -cm}$ along [100] for our results and 
those of Krentsis {\em et al\@.}\cite{krentsis70} respectively.  

To investigate the magnetic phase diagram we performed
field sweeps at constant temperature near the lower transition 
and temperature sweeps at constant field through
both transitions.  The temperature and magnetic field at the
phase boundaries were measured at the point of the resistance
anomaly in the sweep. The set of field sweeps
along [110] near the lower transition is shown in Fig. 3.  
Both increasing and decreasing sweeps
of magnetic field were performed.
The phase transitions are marked by the inflection points of the
resistance curves.  The slight increase in resistivity upon going into the 
commensurate phase may be
due to the formation of a new magnetic superzone, causing extra scattering 
of the charge carriers.  The
resulting magnetic phase diagram is shown in Fig. 4 along with a simple
(quadratic) extrapolation to higher magnetic fields.  We found an identical 
phase diagram
for fields in the [100] direction once again indicating the small anisotropy
in the basal plane.  The phases shown are the only ones present 
in the range of temperature (0.3-300 K) and magnetic field (0-16 T)
accessed by our measurements.  Future experiments, most likely 
specific heat measurements, are planned 
to verify this phase diagram.
In Fig. 4 the N\'{e}el transition line shows a nearly 
vertical increase with field.
However, ${T_{\rm \/ c}}$
noticeably increases with magnetic field along [100] or [110] and has
increased by 6 K in a magnetic 
field of 15 T. When we assume a quadratic dependence of $\vec{H} $
at the transition on 
$T_{\rm \/ c}$ and a slight 
linear change in $T_{\rm \/N}$ we expect
the IC phase to be stable up to a critical field of 
$\mu _{\rm \/ 0} H_{\rm \/ c} \simeq $ 30 T\@.

In spite of the simplicity of the extrapolation involved it is obvious
that the critical field is much larger
than the 1 T predicted by Tarasenko 
{\em et al.}\cite{tarasenko89}  The reason 
for this discrepancy is unclear at this time.  These authors used a 
free energy expansion in powers of the sublattice magnetization combined 
with a Landau-Khalatnikov analysis of the phase stability.
This approach predicts that the incommensurate SDW phase contains longitudinal
and transverse phases which are separated by a first order transition.  
The longitudinal phase would be virtually invisible to experiment
since it only exists over a narrow temperature
range of 20 mK just below the N\'{e}el transition.  This extra phase is
required however if the N\'{e}el transition is second 
order\cite{corliss85} and the C-IC 
transition is first order.\cite{pluznikov82}  
One key approximation in the analysis of Tarasenko {\em et al\@.}
\cite{tarasenko89} 
is that the anisotropy between the in-plane directions is assumed 
to be zero and the anisotropy between the basal plane and c-axis 
is assumed to be infinite.  Inelastic neutron scattering measurements
of the spin waves in the commensurate phase have found that
the anisotropy in the magnetic interactions along [001] and [100] 
is about a factor
of two.\cite{holden94}  Also, estimates had to be made for
the coupling constants between the magnetic field and the 
fluctuations 
which appear in the Landau-Khalatnikov equations of motion.  Hopefully 
the new data as well as
some of the concepts introduced by Corliss {\em et al\@.}\cite{corliss85} 
with regard to the 
phase diagram will contribute towards a refined prediction that agrees with 
experiment.  Such a refinement will be useful in
determining the magnetic properties of related compounds 
with SDW transitions to commensurate or incommensurate phases
such as gallium or arsenic doped ${\rm FeGe_{2}}$ 
and Cr+0.2at\%V\cite{noakes90} and could also give insight into the
interactions at work in the high ${\rm T_{\rm c}}$ oxides.

We have also performed low-temperature (0.3 K) magnetoresistance measurements
in magnetic fields up to 16 T along the [001] and [110]
directions in transverse and longitudinal configurations and also for
current along [100] and magnetic field along [001].  All sets
of transverse measurements show substantial magnetoresistance in
high fields.  The behavior we see is intermediate to closed orbits
and open orbits since there is no saturation of magnetoresistance
as one would expect for a closed orbit and
no sample shows an $H^{\rm \/ {2}}$ dependence 
that would clearly signify an open orbit.
The best description of the data between 1 T and 16 T is given 
by $H^{\alpha }$ with $\alpha  < 2$ in all cases with the basal plane
transverse magnetoresistance giving values of $\alpha$ slightly less
than 2.  These fits for current along [110] and [001] along with 
the exponents, $\alpha$, are shown in Fig. 5.
Other descriptions of the field dependence, such as polynomials, were 
less successful for the same range of magnetic fields and the data shown.
The observed transverse magnetoresistance could arise from a combination of
closed and open orbits complicated by impurities in the
sample.  Bandstructure calculations \cite{grechnev91} show 
a Fermi surface with two sheets and the absence of open orbits in
the [110] and [001] directions with
an open orbit along [100] near the (001) face of the zone boundary.
The gross behaviour for $i \: \parallel$ [100] is
well described by a quadratic
function over the entire field range or a power law with $\alpha $ = 
1.37 for the range from 1 to 16 T.
Another interesting feature of the [001] transverse magnetoresistance
measurements is shown in Fig.
6 where the magnetoresistance at 10 T is higher than 
the resistance in zero field at $70 \: {\rm K}$ and we can clearly see 
the decrease in the magnitude of the magnetoresistance
with increasing temperature.  This result is also suggestive of open orbits
due to the effect of increasing temperature on Fermi surface sharpness.

The most striking feature of the low temperature transverse 
magnetoresistance is the occurrence of Shubnikov-de Haas
oscillations for $\vec{H} \parallel$ [001] and $i \: \parallel$ [100].
We show these oscillations in Fig. 7.  They remain
clearly visible up to 4 K  and for magnetic fields as low as 7 T. 
They can also
be detected in the $i \: \parallel$ [110] and $\vec{H} \parallel$ [001] 
data but are not nearly so prominent.  The frequency of oscillations
is $190 \: \pm \: 10 \: {\rm T}$ corresponding to a Fermi surface extremal
area of $(1.8 \: \pm \: 0.1) \times 10^{-2} \:{\rm \AA} ^{-2}$.  Previous
observations \cite{svechkarev80} had found frequencies
for $\vec{H} \parallel$ [001] of $35 ~ {\rm T}$ and 1630 T.  
We found no evidence for these frequencies but with our field resolution,
the high value of magnetic field where the oscillations appeared, and the
presence of the oscillations at 190 T it
would have been very difficult to observe either of these frequencies. 
It is interesting to note that a circular cross 
section would give a diameter
of  $0.142 \: \pm \: 0.004 ~ {\rm a}$* which is the same order of magnitude
as the degree of discommensuration (0.05 a*) possibly identifying the part
of the Fermi surface responsible for wavevector nesting
in the incommensurate phase.  By carefully examining the oscillations
at various temperatures, $T$, and magnetic fields, $H$, using a model
\cite{shonenberg84} 
for the amplitudes, $A \/$:
\begin{equation}
A \propto T \sqrt{H} ~ \frac{\exp \left(-\frac{2 \pi^{2} m^{*}ck_{\rm B} T'
}{e \hbar H}\right)}{{\rm sinh}\left(\frac{2 \pi^{2} m^{*}ck_{\rm B} T
}{e \hbar H}\right)}
\end{equation}
giving a
Dingle temperature, $T'$, of $1.2 \: \pm \: 0.1 ~ {\rm K}$ and 
an effective 
mass, $m^{*}$, of $0.45 \: \pm \: 0.05 ~ {\rm electron}$ masses.
This effective mass is comparable with previous measurements
\cite{grechnev91} which
gave a value of $0.51 \: \pm \: 0.02 ~ m _{\rm e}$ for the high frequency
oscillation for $\vec{H} \parallel$ [001]. 
The appearance of strong 
oscillations in the transverse magnetoresistance
is another indicator that 
the large magnetoresistance is due to a Fermi surface 
effect rather than the response of the magnetic 
structure to the applied field.  Given the large N\'{e}el temperature and
the large value of the critical field\cite{tarasenko89} it is unlikely
that the magnetic fields used could cause such a substantial change in
the resistivity ($15 ~ \mu \Omega {\rm -cm}$ at 16 T) when at the phase 
transitions themselves we only see a change of less than a percent
($0.25 ~ \mu \Omega{\rm -cm}$). 

In conclusion, we have determined the magnetic phase diagram of the
itinerant SDW antiferromagnet ${\rm {FeGe_{2}}}$ for fields up to 16 T in
the basal plane.
The incommensurate phase that separates the paramagnetic from the
commensurate phase is stable up to $\sim $ 30 T, in disagreement with
predictions based on a Landau theory.
Experimental determination of exchange and anisotropy constants and
perhaps refinement of the approximations in the theory are
necessary to resolve this inconsistency.  Magnetization and
neutron scattering experiments along these lines are 
planned for the near future.
The low-temperature magnetoresistance shows a moderate anisotropy
between the c-axis and the basal plane.  The transverse magnetoresistance
can be qualitatively understood on the basis of the reported bandstructure.
The appearance of a new frequency of quantum oscillations at $190 \pm 10 \:
{\rm T}$ has identified a section of the Fermi surface that
is of the same order of magnitude ($(1.8 \: \pm \: 0.1) 
\times 10^{-2} \:{\rm \AA} ^{-2}$) as that for Fermi surface nesting
of the incommensurate spin density wave.  This provides
an important clue
to the connection between the magnetic structure and electronic structure.
The effective carrier mass is in agreement with previous measurements of
other extremal orbits in this material.
 
\section*{Acknowledgements}
We are grateful to  Prof. A. Z.  Menshikov for providing 
the single crystal samples
which were prepared at the Ural Polytechnical University, and J. M. Perz
and T. M. Holden for helpful discussions.  
This work was supported by the Natural Sciences and Engineering Research 
Council of Canada and the Canadian Institute for Advanced Research.

\begin{figure}
\caption[Structure of FeGe2]{The crystal structure of ${\rm FeGe_{2}}$.
We show the unit cell defined by the Fe atoms ({\small $\bigcirc $}) with
the c-axis in the vertical direction.  We
can see the interceding Ge atoms ($\bullet $) lie in the basal plane 
with a glide transformation connecting the neighboring layers.}
\end{figure}   

\begin{figure}
\caption[Resistivity vs temperature]{Resistivity as a 
function of temperature in zero field from 0.3 K to 300 K.  Measurements 
for current $\parallel $ [100], [110], and [001] are shown.
Notice the large anisotropy between the in-plane and out-of plane
directions: the room temperature 
resistivities differ by a factor of 2.
The inset highlights the phase transitions observed in measurements along
[110].} 
\end{figure}

\begin{figure}
\caption[Magnetic field sweeps for transitions]{Set of magnetic field 
sweeps along [110] near the lower transition 
temperature $T_{\rm \/ c}$, showing the phase boundary.
The temperatures are, from
the bottom to the top, 260 K, 263.6 K, 264 K, 265 K, 266.1 K, 267 K,
267.7 K, 268.5 K, and 269 K. Transitions are shown by arrows.  Notice
the difference in the curve at 260 K where no transition is present.  
The uncertainty in $\mu _{\rm \/ 0}H$ at the transition is  
0.2 T\@.  The curves have been offset for sake of clarity.
The $T$ = 260 K  and the $T$ = 264 K curve have not been shifted. The
other curves have been shifted up or down by $0.25 \: \mu \Omega {\rm -cm}$ 
for each step away from 264 K.}
\end{figure}    

\begin{figure}
\caption[Magnetic phase diagram]{Magnetic phase diagram of 
${\rm FeGe_{2}}$ for ${\vec{H} \parallel}$ [110] constructed from 
data in Fig. 3 and temperature
sweeps that are not shown.  Solid lines are used in our experimental
range up to 16 T.
The dashed lines are a simple extrapolation to higher fields:
$H(T)/H_{\rm \/ c}=1-\left ((T-T_{\rm \/ N}(H_{\rm \/ c}))/
(T_{\rm \/ c}-T_{\rm \/ N}(H_{\rm \/ c}))\right )^{2}$ 
for the C-IC transition
and $H(T)/H_{\rm \/ c}=1-T/(T_{\rm \/ N}(H_{\rm \/ c})-T_{\rm \/ N})$ 
for the N\'{e}el transition where $T_{\rm \/ N}(H_{\rm \/ c})$ is the
N\'{e}el temperature at the critical field.
The phase diagrams for fields along [110]
and [100] are virtually identical.}
\end{figure} 

\begin{figure}
\caption[Fits to magnetoresistance]{Magnetoresistance
$\Delta \rho / \rho  _{\rm \/ H=0}$ data
for current along [110] and [001] for longitudinal and transverse 
magnetic fields.
The solid lines
are power laws $\Delta \rho / \rho \/ _{\rm \/ H=0} \propto H^{\alpha}$ for 
fields greater than 1 T.  
The values of $\alpha $ are, from top to bottom, 1.41(5), 1.54(5),
0.79(6), and 0.48(6).  The uncertainties arise from the
variation in fit over different magnetic field ranges.
Some data points have been omitted for clarity.}
\end{figure}

\begin{figure}
\caption[Temperature sweeps]{Variation in magnetoresistance with 
temperature for $i \parallel$ [001] and ${\vec{H} \parallel}$ [110].
}
\end{figure}
\begin{figure}
\caption[Quantum oscillations]{Resistivity for $i \parallel$ [100] 
and ${\vec{H} \parallel}$ [001] versus magnetic field at 0.3 K.  
The panel on the left shows the raw data and the panel on the right
highlights the quantum oscillations.  The data pictured is the residual
from a quadratic fit plotted versus $1/H$ giving a frequency of $190 \:
\pm \: 10 ~ {\rm T}$ and an effective mass of $0.45 \: \pm \: 0.05 ~ 
m _{\rm e}$.}
\end{figure}  
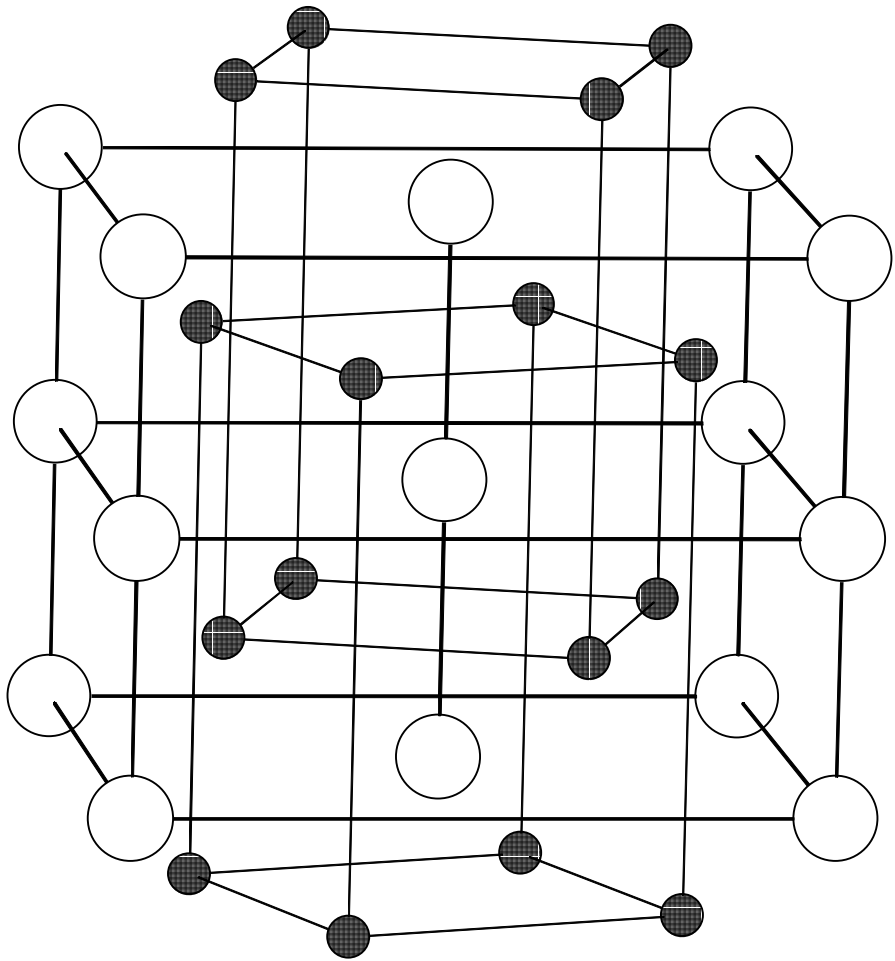
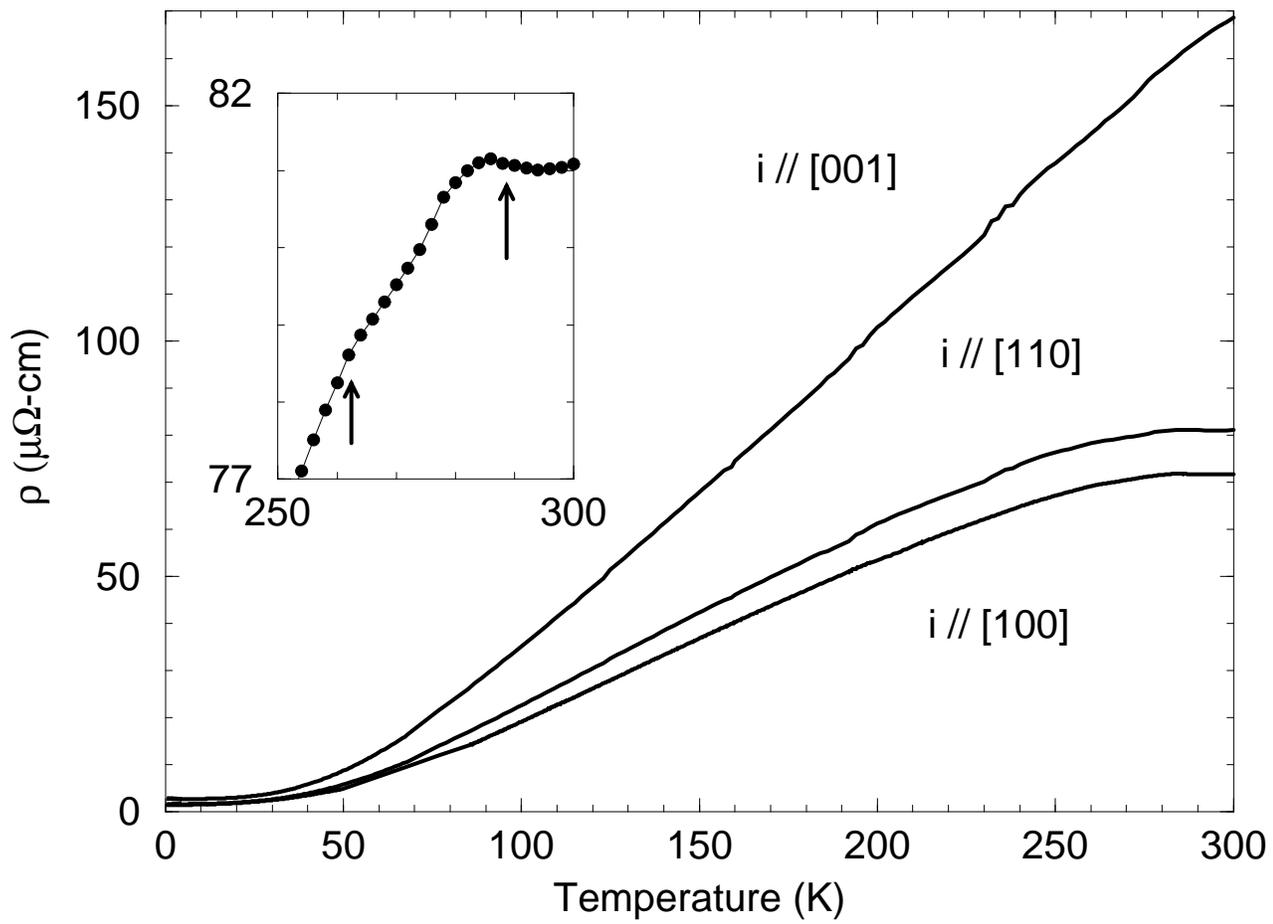
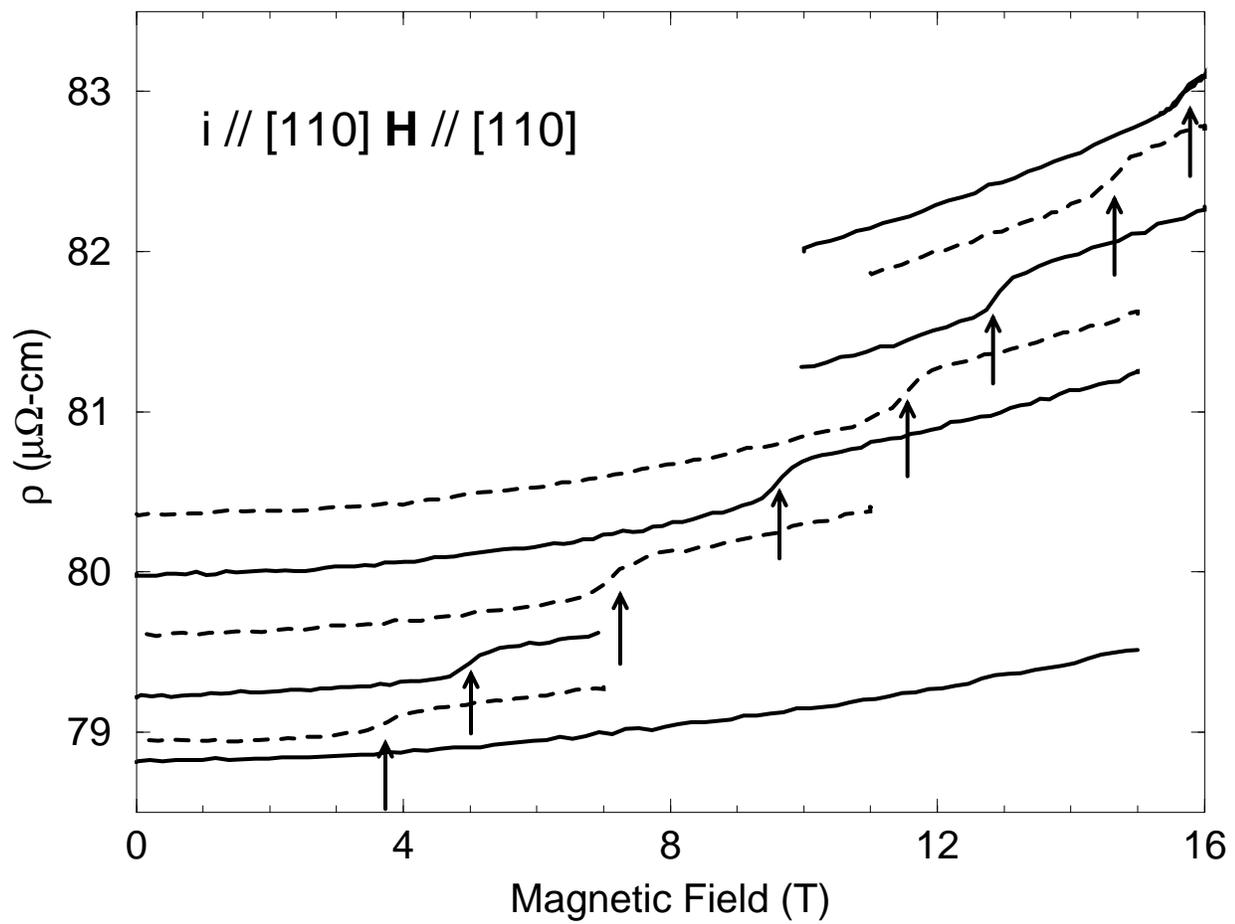
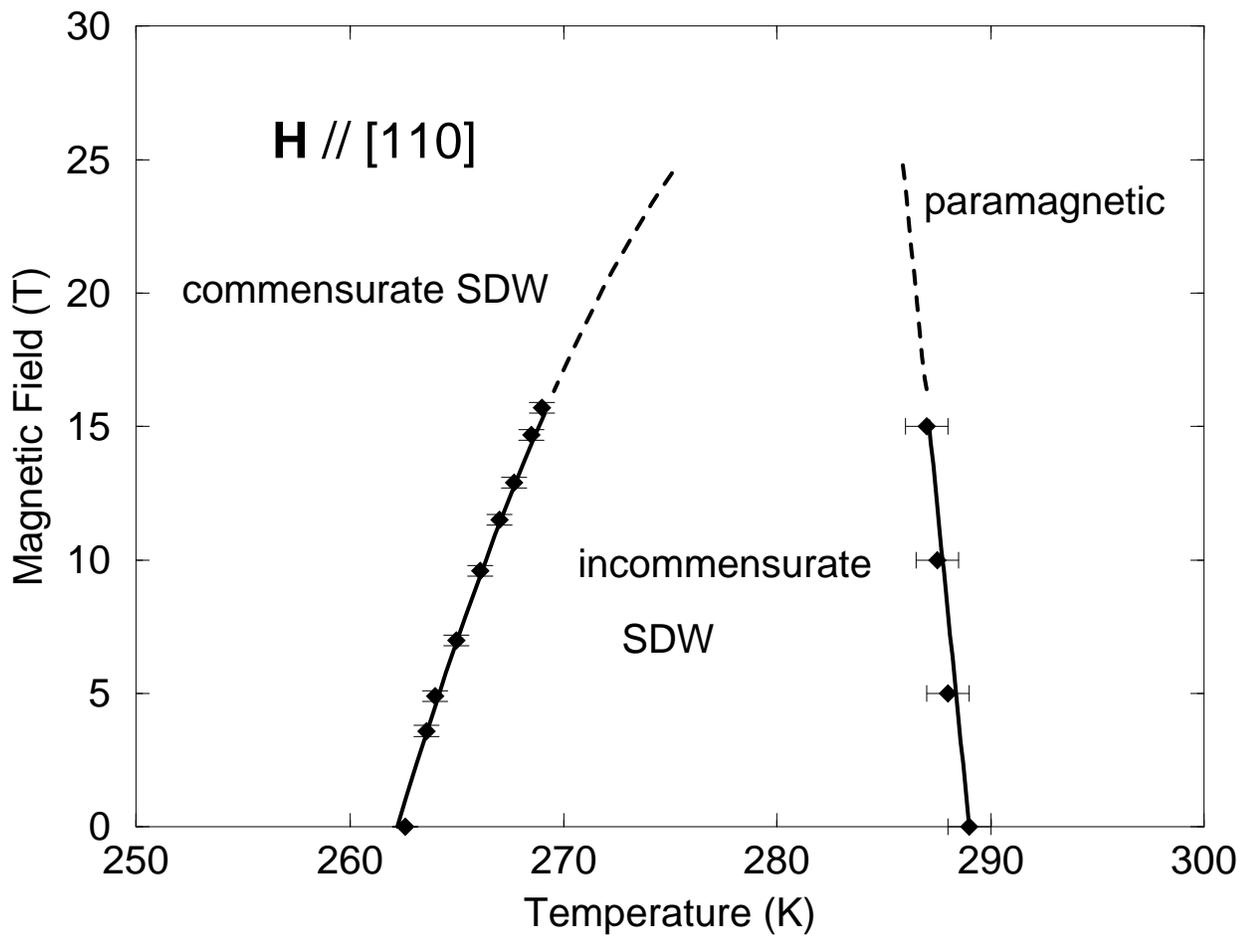
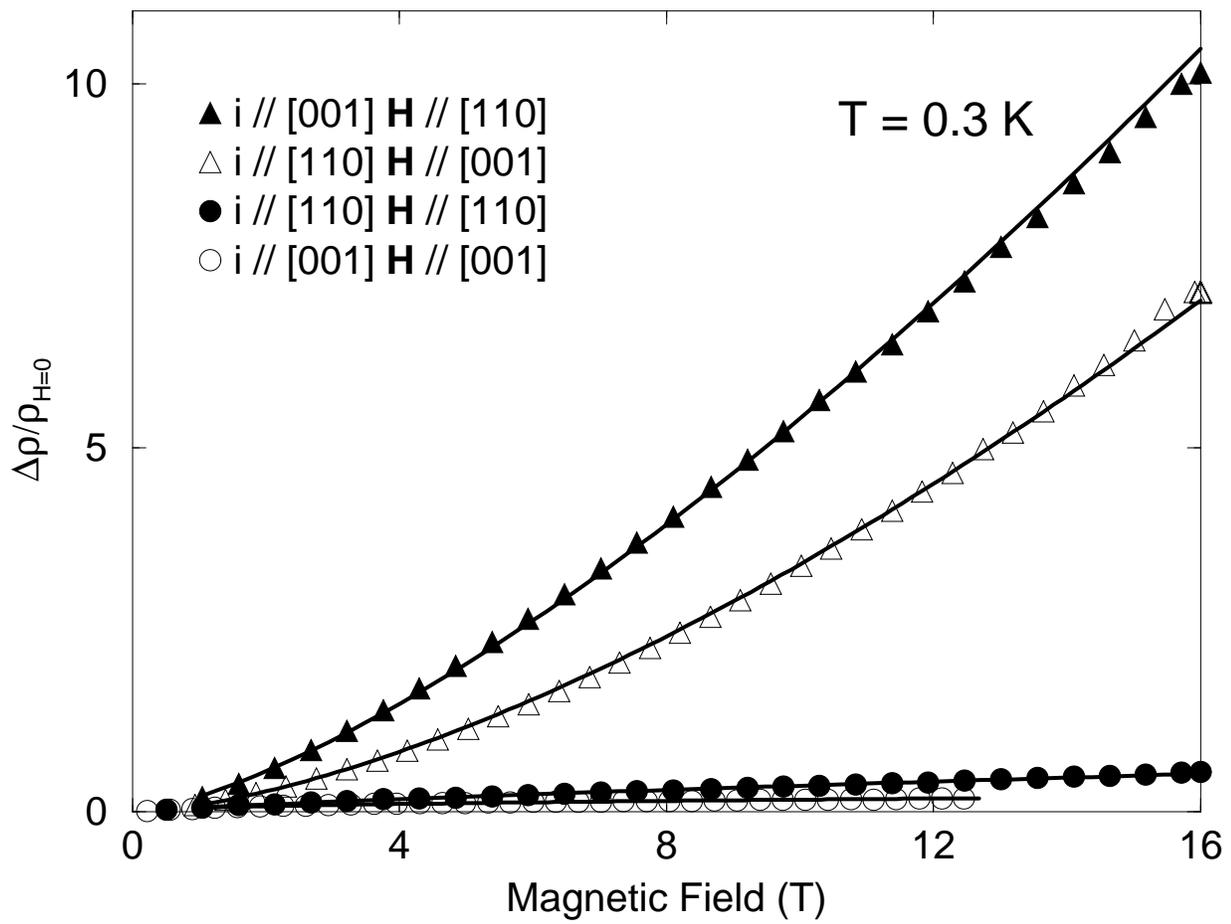
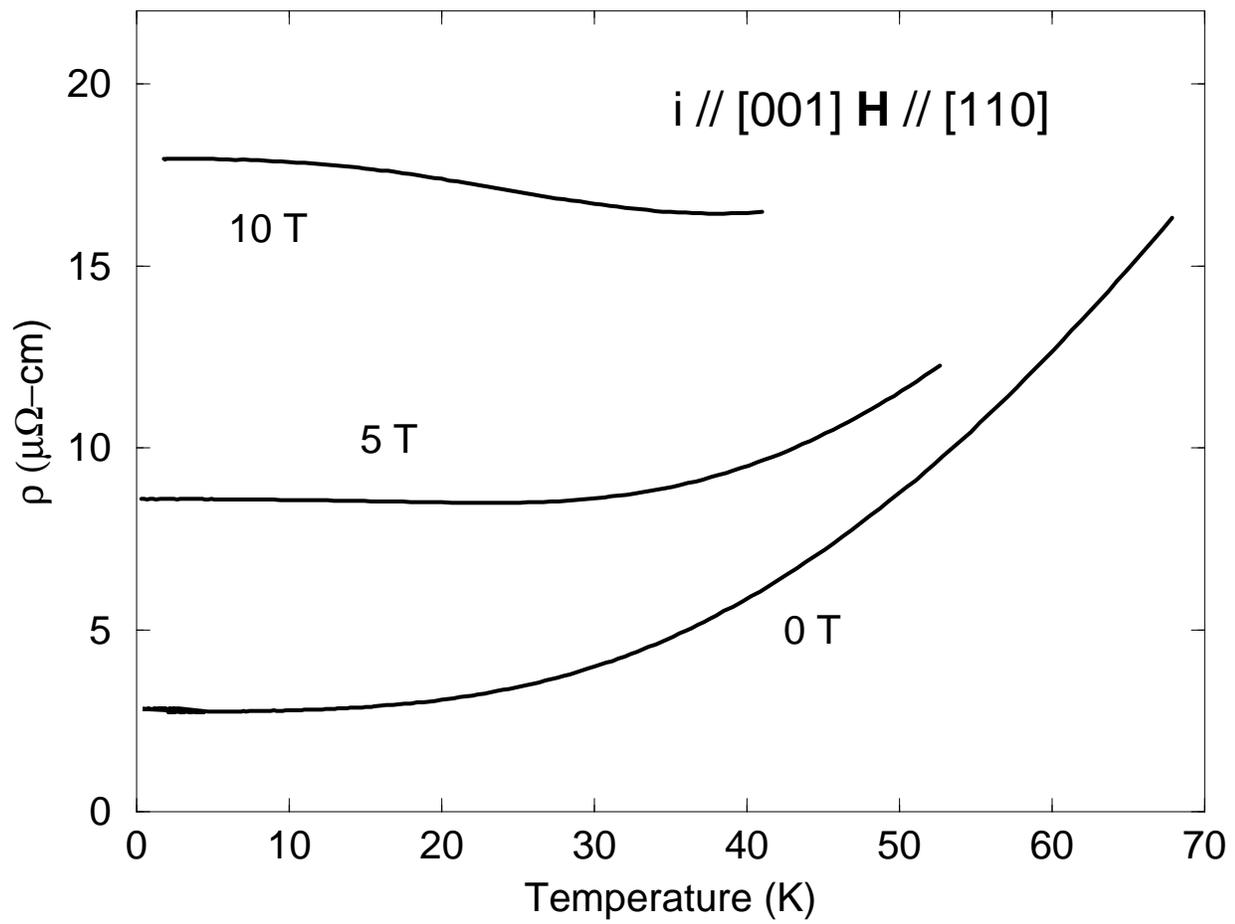
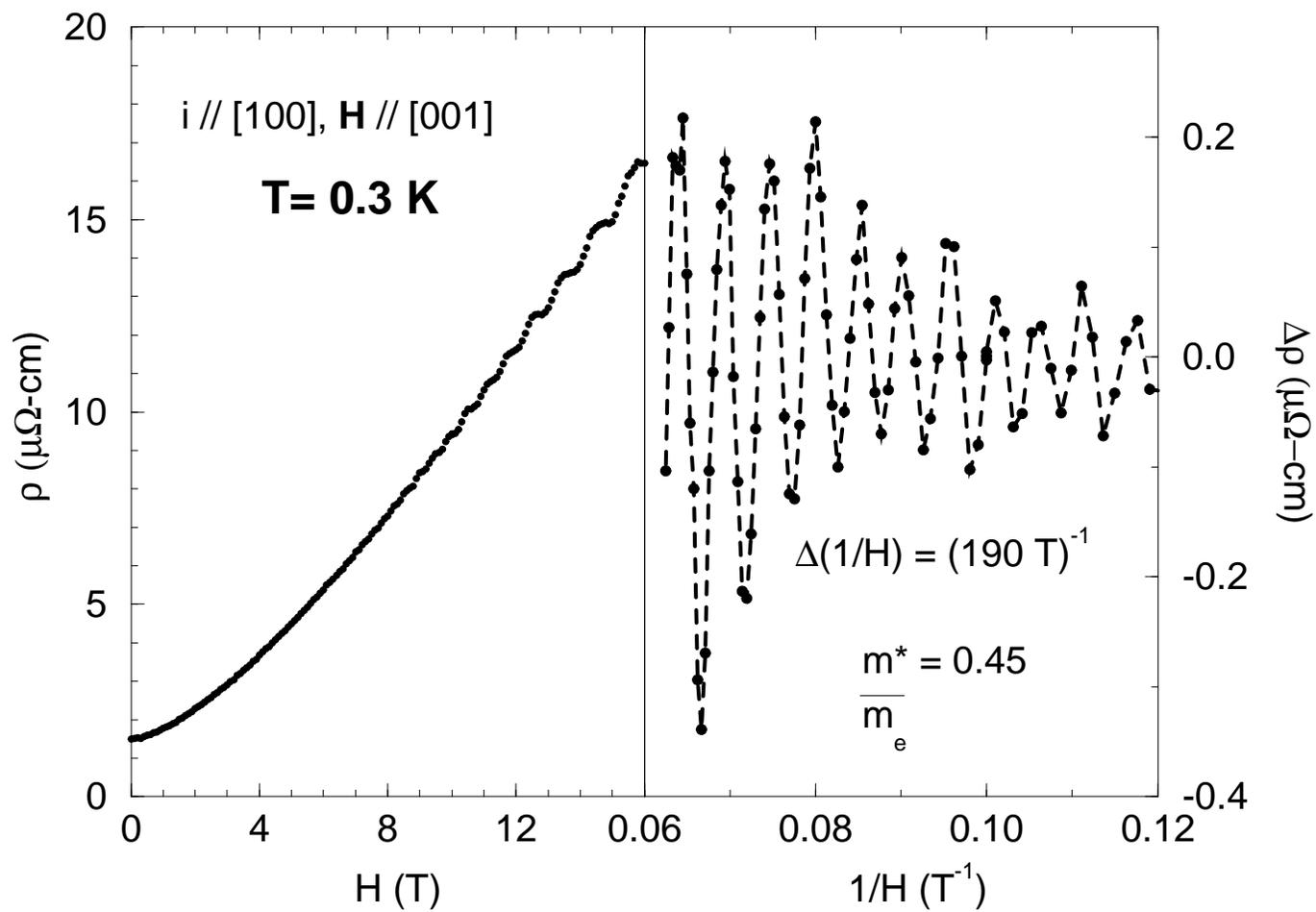
\end{document}